\documentclass[10pt,journal,onecolumn,twoside,draftcls]{IEEEtran}
\usepackage{amsfonts,color}
\usepackage{amssymb,amsmath,mathrsfs}
\usepackage{bm}
\usepackage{mathtools}
\usepackage[noadjust]{cite}
\usepackage[pdftex]{graphicx}
\usepackage{epstopdf}
\usepackage{dblfloatfix} 
\usepackage{float}
\usepackage{relsize}
\usepackage{lipsum}
\title{{\huge Stochastic Geometry Interference Analysis of\\ Radar Network Performance}}
\author{Andrea Munari, Ljiljana Simi\'{c}, Marina Petrova
\thanks{A. Munari and L. Simi\'{c} are with the Institute for Networked Systems of the RWTH Aachen University, D-52072 Aachen, Germany (e-mail: \{firstname.lastname\}@inets.rwth-aachen.de).
M. Petrova is with the School of Electrical Engineering and Computer Science, KTH Royal Institute of Technology, 10044 Stockholm, Sweden (e-mail: petrovam@kth.se).}
\vspace{-1em}}

\newtheorem{prop}{Proposition}

\newcommand{\eq}[1]{\mbox{\ensuremath{#1}}}
\newcommand{\mc}[1]{\ensuremath{\mathcal{#1}}}
\newcommand{\pr}{\ensuremath{\mathbb P}}

\newcommand{\pd}{\ensuremath{\mathsf{P}_{\mathsf{d}}}}
\newcommand{\pfa}{\ensuremath{\mathsf{P}_{\mathsf{fa}}}}
\newcommand{\dm}{\ensuremath{d_{\mathsf{m}}}}

\newcommand{\Fis}{\ensuremath{F_{\mathcal{I}_s}}}
\newcommand{\fis}{\ensuremath{f_{\mathcal{I}_s}}}

\newcommand{\sInt}{\ensuremath{{\mathcal{I}_s}}}

\begin{document}

\maketitle
\thispagestyle{empty} \setcounter{page}{0}

\begin{abstract}
This work characterises the effect of mutual interference in a planar network of pulsed-radar devices. Using stochastic geometry tools and a strongest interferer approximation, we derive simple closed-form expressions that pinpoint the role played by key system parameters on radar detection range and false alarm rate in the interference-limited region. The fundamental tradeoffs of the system between radar performance, network density and antenna directivity are captured for different path-loss exponents in the no-fading and Rayleigh-fading cases. The discussion highlights practical design hints for tuning the radar parameters. The accuracy of the model is verified through network simulations, and the role of random noise on detection in sparse, non interference-limited networks is characterised.  
\end{abstract}


\section{Introduction} \label{sec:intro}

Compact low-cost radar devices are set to become pervasive for providing short-range environmental awareness in emerging applications such as automotive~\cite{Kandeepan17_ITS}, enhanced localization~\cite{Guidi16_TMC}, or radio resource optimization~\cite{Simic16_RadCom}. In future heterogeneous networks, low-power radars are also envisioned to share spectrum -- and possibly be co-located -- with communication devices~\cite{Guidi16_TMC, Heath18_TVT}, in e.g. the $60$~GHz unlicensed band. These scenarios give rise to coexistence of a multitude of radar devices, randomly oriented over a large area, sharing a frequency band in an uncoordinated fashion. For such radar networks, it is paramount to understand the effect of mutual interference on the achievable detection and false alarm rates.

Mutual radar interference has been thoroughly studied in simple two-node topologies~\cite{Brooker07_TEC}. Recent research has started to address more general configurations. The first results were reported in~\cite{Jondral13_EURASIP}, focusing on OFDM radars. The achievable detection probability for different network densities was investigated, relying on a Gaussian approximation of the aggregate interference. A step forward was taken in~\cite{Kandeepan17_ITS}, with the introduction of a stochastic geometry framework to study the performance of a linear automotive radar network. The authors considered an SIR-based detection model, and derived results under different statistical distributions for the radar positions, only for the no-fading case, with a path-loss exponent of $2$, and effectively omni-directional antennas. Random access for the radars was also proposed to mitigate interference.

However, the fundamental performance limits in terms of detection range for a radar immersed in a planar field of interferers are not yet fully characterised. In this work, we propose a simple analytical approach to study a network of independent pulsed-radar devices. Exploiting a strongest-interferer approximation, we derive compact closed-form expressions for the radar detection performance, both for the no-fading and Rayleigh-fading cases and for any path-loss exponent. Our model clearly and comprehensively captures the effect of key system parameters. We study the tradeoff between a desired detection/false-alarm performance and the radar network density and antenna directivity, yielding practical insight into system design rules for tuning the radar network.

\section{System Model and Preliminaries} \label{sec:sysModel}

We consider a population of pulsed-radar devices, modelling their locations as a homogeneous Poisson point process (PPP) $\Phi = \{ \mathbf{x}_i\}$ of intensity $\lambda$ over $\mathbb R^2$. 
Time is divided in slots of equal duration, and devices continuously operate with a common pulse repetition frequency $\delta = 1/M$ [slot$^{-1}$]. Namely, each device follows cycles of duration $M$, transmitting a pulse over one slot, and then waiting for a target echo over the subsequent $M-1$ slots.\footnote{For a bandwidth $B$, a slot duration $1/B$ can be assumed;  $B$ influences the radar resolution, investigating which is out of scope of this work.} While slot-level synchronisation eases system modelling, the uncoordinated nature of the radar population is captured by allowing random offsets among the operating cycles of different nodes. Accordingly, each PPP element is assigned an independent mark \mbox{$m_i \sim \mathcal U \{0,M-1\}$}, and transmits its pulses at slots $m_i + kM$, $k\in \mathbb N$.

Within this setup, we are interested in characterising the impact of radar-to-radar interference on the detection performance. We focus without loss of generality on the \emph{typical} node located at the origin of the plane, and consider a power-based detection rule. Specifically, a target is declared present if the aggregate incoming power over any of the $M-1$ slots spent listening for the echo exceeds a threshold $\Theta$. All radars transmit with power $P_t$, and have a planar antenna pattern with maximum gain $\mathcal G_m$, where the boresight direction of each device is modelled as an independent and uniform r.v. in $[0,2\pi)$.  Signal propagation undergoes path-loss with exponent $\alpha > 2$, and we investigate both the no-fading and  Rayleigh-fading cases. For a target of radar cross section $\sigma$ in the boresight direction of the typical node, the incoming reflected power $\mathcal S$ follows from the well-known range equation as
\begin{equation}
\mathcal S = \frac{P_t \, \mathcal G_m^2  \, \kappa \,\sigma \,\ell}{4\pi} \cdot \zeta \,d^{-2\alpha} \,.
\label{eq:echoPower}
\end{equation}
Here, $\ell := [\,c/(4\pi f)\,]^2$, $c$ is the speed of light, $f$ is the carrier frequency, $\kappa$ is a signal processing gain, $d$ is the target distance, and $\zeta \sim \exp(1)$ in the case of Rayleigh fading or is set to $1$ otherwise. By the properties of thinning for PPP, the aggregate interference $\mathcal I$ the typical radar experiences over each of the $M-1$ detection slots is i.i.d., taking the form
\begin{equation}
\mathcal I = \sum\nolimits_{\mathbf x \in \Phi_{\mathcal I}} P_t  \,\mathcal G_t \mathcal G_r  \, \ell \, \zeta_{\mathbf x}  \, \Vert \mathbf x \Vert^{-\alpha}.
\label{eq:aggIntDef}
\end{equation}
In \eqref{eq:aggIntDef}, $\Phi_{\mathcal I}$ is a thinned version of $\Phi$ with intensity \eq{\lambda' = \delta \lambda} capturing the active interferers over the observed slot. In turn, $\mathcal G_t$ and $\mathcal G_r$ are the transmit and receive antenna gains for the link between the radar transmitter located at $\mathbf x$ and the typical receiver (thus depending on the random boresight orientation of the radars). Finally, for the same link, $\zeta_{\mathbf x}$ is once more either an exponential r.v. of unit mean in the case of Rayleigh fading, or is set to $1$ when no fading is considered. 

For the introduced system model, a target is in general correctly detected with detection probability \pd, expressed as $\pr\left\{ \mc{S} + \mc{I} + W \geq \Theta \right\}$, where $W$ is a random variable accounting for noise. Given the interference-limited nature of the networks under consideration, discussed in more depth in App.~\ref{app:noise}, we will however assume in the remainder $\mathcal I \gg W$, and disregard the noise component.
Following this approach, a \emph{false alarm} is triggered when the power over at least one of the observed slots exceeds the detection threshold in the absence of the desired echo, i.e. with probability \mbox{$\pfa = 1 - F_{\mathcal I}(\Theta)^{M-1}$}, where $F_{\mathcal I}(i) = \pr\left\{ \mc{I} \leq i \right\}$ is the cumulative distribution function of $\mathcal I$. Following common practice in radar design, we will set $\Theta$ so as to achieve a tolerable false alarm rate \pfa, and evaluate the performance of the system in terms of the corresponding detection probability. 

\section{Radar Interference Statistics} \label{sec:analysis}

As highlighted in Sec.~\ref{sec:sysModel}, a statistical description of the aggregate interference is required to properly set the radar detection threshold. Considering omni-directional antennas, classical stochastic geometry results characterise $\mathcal I$ as L\`{e}vy distributed for $\alpha=4$, while no closed-form solution is known for other path-loss exponents. In order to derive compact and insightful expressions for a broader range of parameters,  we follow a different approach, tuning $\Theta$ based on the power statistics of the strongest interferer only. For the sake of tractability, we furthermore initially focus on an ideal cone antenna pattern, assuming the gain to be $\mathcal G_m$ over a beamwidth $\varphi$ and zero elsewhere,\footnote{For any beamwidth, the gain is set as $\mathcal G_m = 4\pi/\varphi^2$. 
The impact of more realistic antenna patterns as well as the tightness of the strongest interferer approximation will be discussed in detail in Sec.~\ref{sec:results}.} to obtain:
\begin{prop}
Let \sInt\ be the power of the strongest interferer at the typical node over a detection slot. Then, under the cone antenna pattern assumption, we have
\begin{equation}
\Fis(i)  := \pr\{\sInt \leq i \} = \exp\left(  -\frac{\lambda \,\delta \,\varphi^2\, \Omega \,\omega^{2/\alpha}}{4\pi} \cdot i^{-2/\alpha}\right)
\label{eq:cdfInt}
\end{equation}
\label{prop:cdf}
\end{prop}
where $\omega := P_t \,\mathcal G_m^2 \ell$, $\Omega=1$ for the no-fading case and $\Omega = \Gamma(1+2/\alpha)$ in the presence of Rayleigh fading, with $\Gamma(x)= \int_{0}^{\infty}x^{t-1} e^{-x}\,dt$. \\
\begin{proof}
For the considered model, a transmitter active over the observed slot interferes at the typical receiver only if the alignment of their two randomly oriented antenna patterns overlaps. By simple geometrical arguments, this event has probability $\varphi^2/(4\pi^2)$. By virtue of this further thinning, the aggregate interference is driven by the homogeneous PPP $\Phi_{\mathcal I}'$ of intensity $\delta \lambda \varphi^2/(4 \pi^2)$. 
Following a reasoning similar to \cite{Haenggi14_TIT}, let us introduce a new PPP \mbox{$\Xi := \{ \xi_i = (\,\left\Vert \mathbf x_i \right\Vert^{-\alpha} \zeta_{\mathbf x_i} )^{-1}, \mathbf x_i \in \Phi_{\mathcal I}' \}$} over $\mathbb R^+$, whose elements are ordered s.t. $\xi_i < \xi_j$, $\forall \,i<j$. From the mapping theorem, $\Xi$ is a non-homogeneous process of intensity measure $\Lambda([0,r]) = \delta \lambda \varphi^2 r^{2/\alpha}\, \Omega /(4 \pi)$, \mbox{$\forall \, r >0$}, where  \mbox{$\Omega := \mathbb E[\zeta_{\mathbf x}^{2/\alpha}]$} evaluates to $1$ with no-fading (i.e. $\zeta_{\mathbf x} \equiv 1$) and to \mbox{$\int_{0}^{\infty} t^{2\alpha} e^{-t} dt = \Gamma(1+2/\alpha)$} with Rayleigh fading (i.e. $\zeta_{\mathbf x}\sim \exp(1)$). Let us indicate as $\mathbf x_{s} \in \Phi_{\mathcal I}'$ the coordinates of the strongest interferer at the typical receiver over the observed slot. Recalling the definition of $\omega$, we get \mbox{$\Fis(i) = \pr\{ (\left \Vert  \mathbf x_s \right\Vert^{-\alpha} \zeta_{\mathbf x_s} )^{-1} \geq \omega / i \}$}. Noting that $(\left \Vert  \mathbf x_s \right\Vert^{-\alpha} \zeta_{\mathbf x_s} )^{-1} = \xi_1$, $\Fis(i)$ can be computed as the probability of not having any point of the process $\Xi$ in the interval $[0,\omega/i)$, i.e. $\Fis(i) = \exp(-\Lambda([0,\omega/i)))$, giving \eqref{eq:cdfInt}.
\end{proof}

Approximating the aggregate interference experienced by a radar receiver with its strongest component, Prop.~\ref{prop:cdf} allows us to derive a compact formulation of the detection threshold needed to achieve a target false alarm probability.  Namely, solving \mbox{$\pfa = 1 - \Fis(\Theta)^{M-1}$} we get
\begin{equation}
\Theta = \omega \, \left( \frac{-\Omega \,(1-\delta) \,\lambda \,\varphi^2}{4\pi \ln(1-\pfa)}\right)^{\alpha/2}.
\label{eq:threshold}
\end{equation}

The simple closed-form expression in \eqref{eq:threshold} provides guidance on how to tune the radar receiver for any path-loss exponent \eq{\alpha>2} as well as for both the Rayleigh-fading and the no-fading case, pinpointing the impact of all relevant system parameters. Remarkably, the presence of fading on the interfering signals is embedded through the sole scaling factor $\Omega$. 

\section{Radar Performance Evaluation} \label{sec:results}

The derived interference statistics allow us not only to properly tune the threshold $\Theta$, but also to evaluate the impact of radar-to-radar interference on the detection performance. To better stress the key tradeoffs for the system under study, we focus on two use-cases of practical interest. Firstly, we consider a setting where links are line-of-sight ($\alpha\simeq 2$) and not affected by fading, representative of mm-wave propagation. We then complement our discussion delving into a scenario characterised by larger path-loss exponents and Rayleigh fading for both radar echo and interference, drawing conclusions that are applicable to traditional cellular and WiFi frequency bands. 
A radar cross section $\sigma=10 \,\rm{m}^2$ and a processing gain $\kappa=10$ are assumed for the target, while devices transmit with power $P_t=10\,\rm{dBm}$ and pulse repetition frequency $\delta=10^{-2}$. Unless otherwise specified, $\pfa=0.1$.

\begin{figure}
\centering
\includegraphics[width=.5\columnwidth]{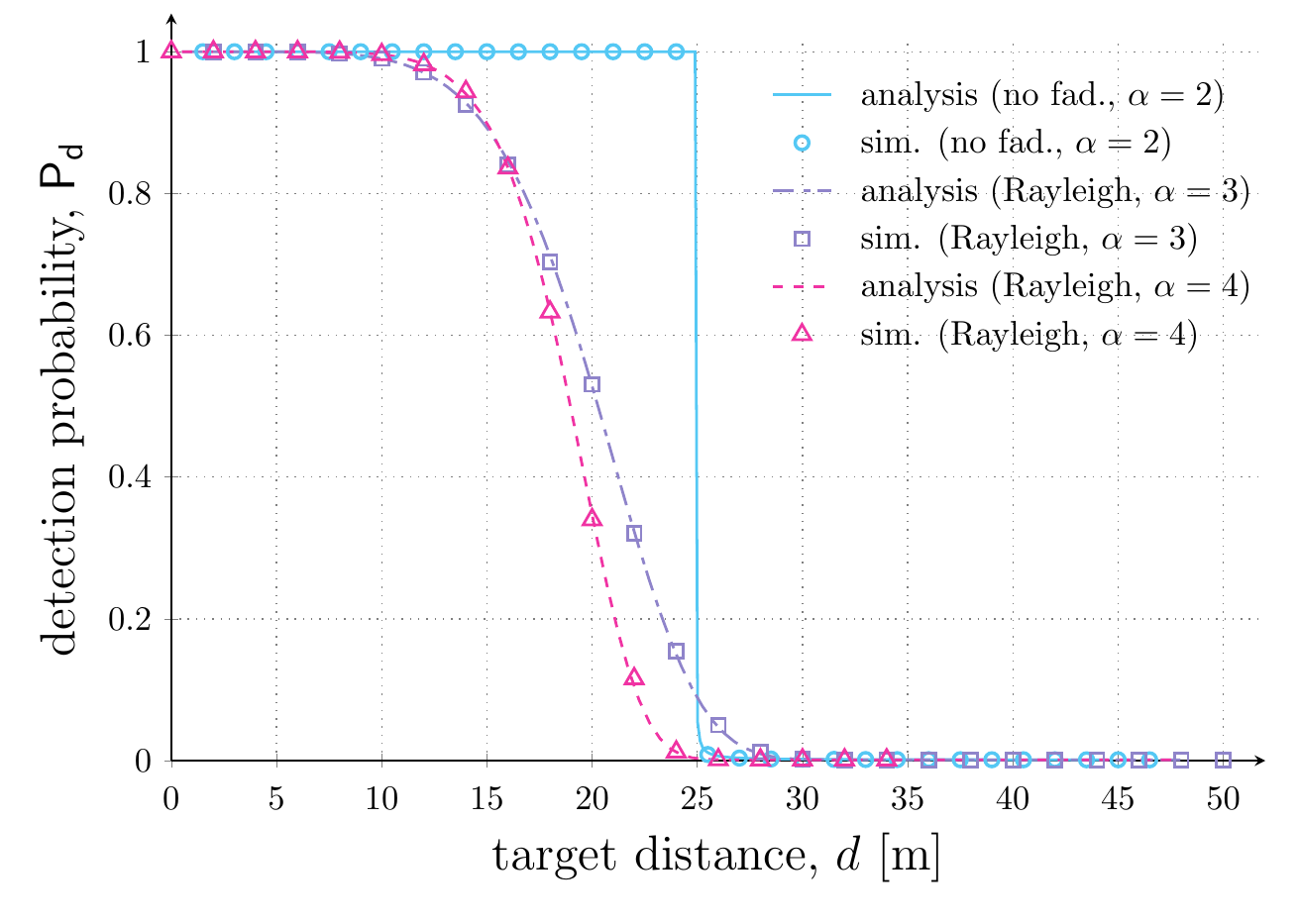}
\caption{\pd\ vs. distance in the no-fading (\mbox{$\alpha=2$}, \mbox{$f=60 \,\rm{GHz}$} -- Sec.~\ref{sec:noFad}) and Rayleigh-fading case ($\alpha=3$, $\alpha=4$, $f=2.4 \,\rm{GHz}$ --Sec.~\ref{sec:fading}). Lines report analytical results (strongest interferer approximation), while markers simulation outcomes considering the aggregate interference. In all setups, $\varphi = \pi/6$, $\lambda=10^{-4}$.}
\label{fig:pdVsDistance}
\vspace{-1em}
\end{figure}

\vspace{-3mm}
\subsection{No-Fading Case}\label{sec:noFad}
In the absence of fading, the presented framework offers a closed-form expression of the target detection probability. Relying on the strongest interferer approximation, we readily get $\pd = 1 - \Fis(\Theta - S)$, where $\Fis(i)$ and $\Theta$ follow from \eqref{eq:cdfInt} and \eqref{eq:threshold} setting $\Omega=1$. The trends obtained for $\alpha = 2$,\footnote{Strictly speaking, analytical results are reported for $\alpha\rightarrow 2$.} and $f=60\,\rm{GHz}$ are presented by the solid line in Fig.~\ref{fig:pdVsDistance}, assuming a  beamwidth $\varphi=\pi/6$ and a density $\lambda=10^{-4}$ [radar/m$^2$].  Given the deterministic nature of the incoming echo power, a target is detected with probability $1$ as long as it is close enough to satisfy the condition $\mathcal S > \Theta$. Conversely, when the sole target reflection 
is not sufficient to exceed the detection threshold, the plot highlights a sharp drop of \pd. This behaviour stems from the strong attenuation undergone by the radar echo, which follows a path-loss power law of exponent $2\alpha$. For $\mathcal S\rightarrow 0$, a detection occurs with probability \eq{\pr\{\mathcal I>\Theta\}=1-(1-\pfa)^{\delta/(1-\delta) } }, which is the asymptotic value the solid curve in Fig.~\ref{fig:pdVsDistance} converges to.

In order to verify the accuracy of the strongest interferer approximation, dedicated simulations were performed. Specifically, multiple PPP instances were generated, so as to extract the statistics of the \emph{aggregate} interference at the typical node and thus compute the detection threshold matching the desired false alarm rate. The probability of target detection was then extracted accounting once more for the whole interference affecting the receiver. The outcome of the simulation study is shown by circle markers in Fig.~\ref{fig:pdVsDistance}. The results show a very tight match with the analytical prediction, confirming that the behaviour of the system is indeed driven by the disruptive effect of having an interferer close to the detecting radar, and further supporting the proposed framework as a simple tool to accurately tune the detection threshold and predict the achievable radar performance. Along this line of reasoning, the trend exhibited by \pd\ pinpoints the existence of a critical distance \dm, separating accurate and missed detection. Setting  \eq{\mathcal S =\Theta}, we readily get
\begin{equation}
\dm = \left( \frac{\kappa \sigma}{4\pi}\right)^{\frac{1}{2\alpha}} \left(\frac{-4\pi \ln(1-\pfa)}{(1-\delta)\lambda \,\varphi^2} \right)^{\frac{1}{4}}.
\label{eq:dm}
\end{equation}
The expression offers interesting insights. Firstly, \eqref{eq:dm} clarifies that the maximum detectable range does not depend on the operating frequency and radar transmission power.\footnote{For very low $\lambda$, i.e. when interference no longer plays a role, detection performance eventually becomes limited by noise power, as discussed in App.~\ref{app:noise}. We also assume $\delta$ to be picked so that the detection range is limited by incoming power rather than by the unambiguous range.} Secondly, the limited impact on \dm\ of the pulse repetition frequency $\delta$ emerges. This insight is non-trivial, as it settles a critical tradeoff. Lower values of $\delta$ potentially favour correct detection of farther targets, both due to the longer echo waiting time and to the weaker interference expected over a single reception slot -- driven by a thinned PPP of intensity $\delta\lambda [\varphi^2/(4\pi^2)]$. Conversely, reducing the pulse repetition frequency calls for a higher $\Theta$ (i.e. a stronger echo for detection) to grant a desired \pfa, since the greater number of slots spent listening favour the occurrence of a false alarm event. The two effects eventually balance each other out, contributing to \dm\ through a scaling factor \eq{(1-\delta)^{-1/4}}, which is negligible for practical values of $\delta$. Finally, the result emphasises the minor role of the path-loss exponent $\alpha$ in determining \dm, unless large radar cross sections come into play.
\begin{figure}
\centering
\includegraphics[width=.5\columnwidth]{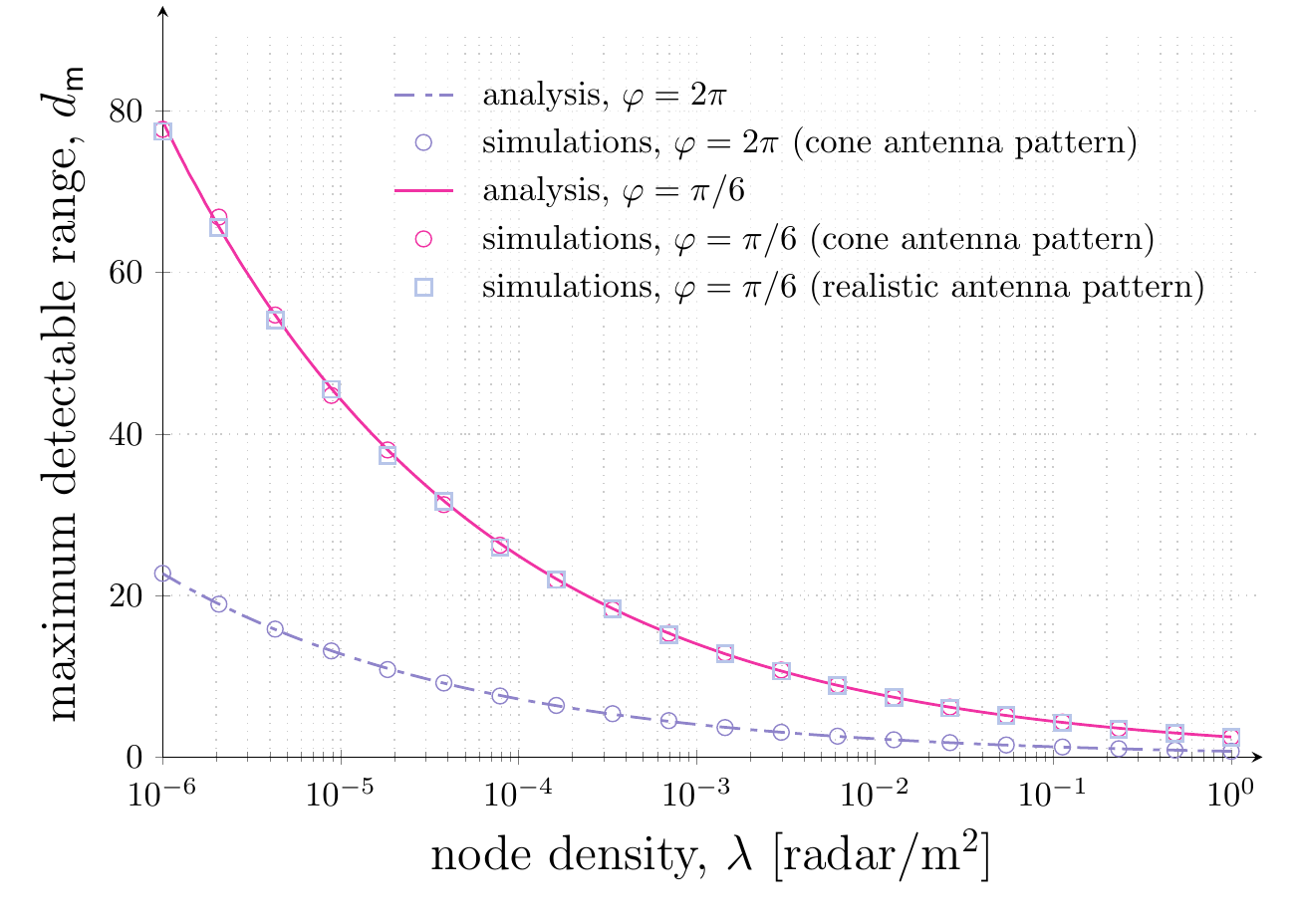}
\caption{\dm\ vs. radar density, no-fading case (\mbox{$\alpha=2$}, \mbox{$f=60 \,\rm{GHz}$}). For the cone antenna model,  lines show analytical results, while circle markers their verification via simulations. Square markers show simulation results for a realistic antenna patterns, $\varphi=\pi/6$.}
\label{fig:dmVsLambda}
\vspace{-1em}
\end{figure}

The maximum detectable distance computed via \eqref{eq:dm} for different beamwidths is shown by solid and dashed lines against $\lambda$ in Fig.~\ref{fig:dmVsLambda}. The accuracy of the analytical approach is again confirmed for all the considered setups by the circle markers, which show the results of system simulations accounting for aggregate interference. The plot clearly highlights the paramount role played by the density of radars sharing the same channel, stressing that for large values of $\lambda$ the detection performance deteriorates sharply. The strong impact of antenna directionality is also apparent. With omni-directional nodes, interference generated by close-by transmitters is disruptive, and targets farther than a few meters can be detected reliably only in very sparse networks. Conversely, narrower beams beneficially filter out undesired signals, enabling detection for \dm\ and $\lambda$ of practical interest already for $\varphi=\pi/6$. 

Two further points follow from the observed trends. Firstly, the considered idealised cone antenna model neglects potentially harmful energy transmitted and received from secondary lobes. To gauge the influence of this, we performed dedicated simulations employing realistic antenna patterns. Specifically, we used the \textsc{Matlab} \emph{Phased Array System Toolbox} to generate a pattern with half-power beamwidth of approximately $25^{\circ}$ and main sidelobes with $10\, \rm{dB}$ attenuation, corresponding to a $4\times 4$ uniform quadratic array of isotropic elements with half-wavelength spacing. The achievable \dm\ is presented in Fig.~\ref{fig:dmVsLambda} with square markers, showing an extremely close match to the analytical results.

\begin{figure}
\centering
\includegraphics[width=.5\columnwidth]{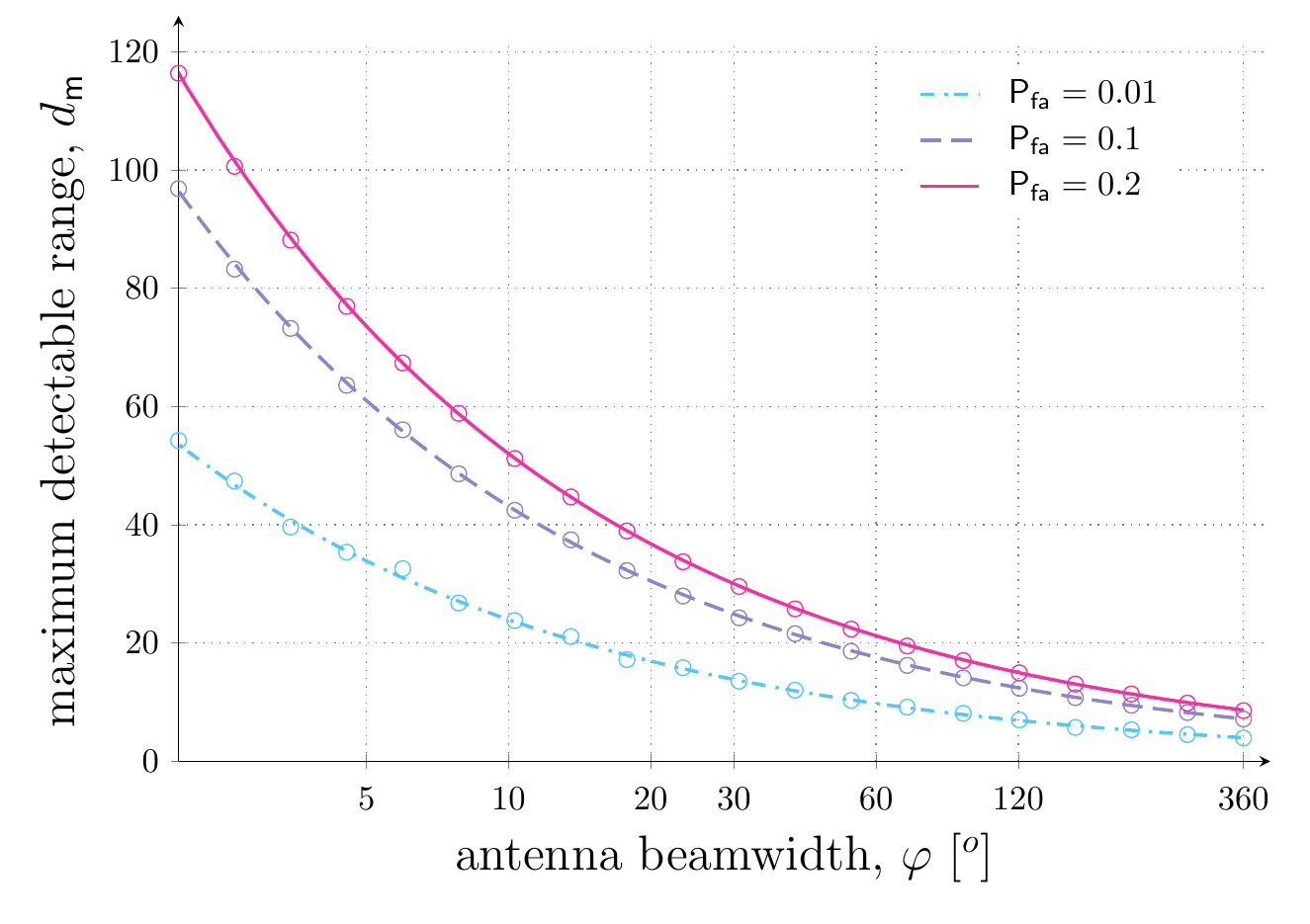}
\caption{\dm\ vs. antenna beamwidth, no-fading case (\mbox{$\alpha=2$}, \mbox{$f=60 \,\rm{GHz}$}, \eq{\lambda=10^{-4}}). A cone antenna model is assumed. Lines show analytical results, while markers simulation results.}
\label{fig:dmVsBeamwidth}
\vspace{-1em}
\end{figure}

Secondly, the need to tune the antenna beamwidth to achieve a desired \dm\ emerges from Fig.~\ref{fig:dmVsLambda} as a critical design choice. This aspect is explored in Fig.~\ref{fig:dmVsBeamwidth}, which depicts the maximum detectable target distance against $\varphi$ for a reference density $\lambda=10^{-4}$ [radar/m$^2$]. Notably, for narrow beamwidths, ranges in the order of $100\,\rm{m}$ are achievable even when immersed in a field of interferers. Finally, the plot also examines the impact of the targeted false alarm probability, reporting \dm\ for three distinct values of \pfa. The study confirms that more stringent requirements in terms of erroneous detection come at the expense of a severe reduction in the detectable range, stressing even more the need for highly directional radar operations.

\vspace{-3mm}
\subsection{Rayleigh-Fading Case} \label{sec:fading}
We extend our study to a setup in which signals undergo multiple reflections caused by the environment.
As a worst case of the multipath-induced  random fluctuations in the received power, we consider a higher path-loss exponent and  Rayleigh fading on the interfering signals and the target echo.\footnote{We note that realistic multipath-induced signal fluctuations would result in performance bounded by the  no-fading and Rayleigh-fading cases.} 
In this case, the detection probability can be computed conditioning on the interference level, i.e. \eq{\pd= \mathbb E_{\mathcal I}[\pr\{\zeta \geq (\Theta -\mathcal I) \,4\pi d^{2\alpha} /(\omega \kappa \sigma) \,|\, \mathcal I\}]}, with \eq{\zeta\sim\exp(1)}. Under the strongest interferer approximation and for a cone antenna pattern we thus have
\begin{equation}
\pd = 1 - \Fis(\Theta) + \int_{0}^{\Theta} \!\!e^{\mathlarger{-\frac{(\Theta-i) 4\pi d^{2\alpha}}{\omega \kappa \sigma}}} \, \fis(i) \,di
\label{eq:pdFading}
\end{equation}
where $\fis(i) := d\Fis/di$ follows from \eqref{eq:cdfInt}, whereas the term $1-\Fis(\Theta)$ accounts for the event of having an interference 
level above the threshold, which marks a target as present regardless of the actual echo power $\mathcal S$. 
The detection rate can then be computed by means of numerical integration for any configuration of the system parameters. Moreover, inspection of \eqref{eq:pdFading} reveals that the analytical expression derived for \pd\ is independent of the transmission power and carrier frequency also in the presence of fading. Namely, $\Theta$ and $\mathcal I_s$ are proportional to $\omega$, which embeds the effect of $P_t$ and $f$ and cancels out in the integrand function (\emph{c.f.} \eqref{eq:threshold}). 

The achievable detection probability for different target distances is depicted by dashed lines in Fig.~\ref{fig:pdVsDistance} for $\varphi = \pi/6$ and path-loss exponents $\alpha=3$ and $4$. Analytical trends are verified also in this case by means of simulations accounting for the level of aggregate interference both in the threshold tuning and the performance evaluation, the results of which are shown by square and triangle markers. As expected, the step transition between accurate and poor detection discussed for the no-fading case is replaced by a smooth degradation, caused by the statistical fluctuations of the echo power. On the other hand, Fig.~\ref{fig:pdVsDistance} confirms the small impact of varying the path-loss exponent when aiming for reliable detection  also in the presence of random signal variations. Indeed, for sufficiently high values of \pd\ (e.g. $\pd\geq0.9$), the beneficially lower level of interference brought by harsher propagation conditions is counterbalanced by the weaker incoming echo. In turn, the latter effect prevails for larger target distances due to the path-loss power law of exponent $2\alpha$ for the useful reflection, leading to the sharper performance decay observed for $\alpha=4$.

\begin{figure}
\centering
\includegraphics[width=.5\columnwidth]{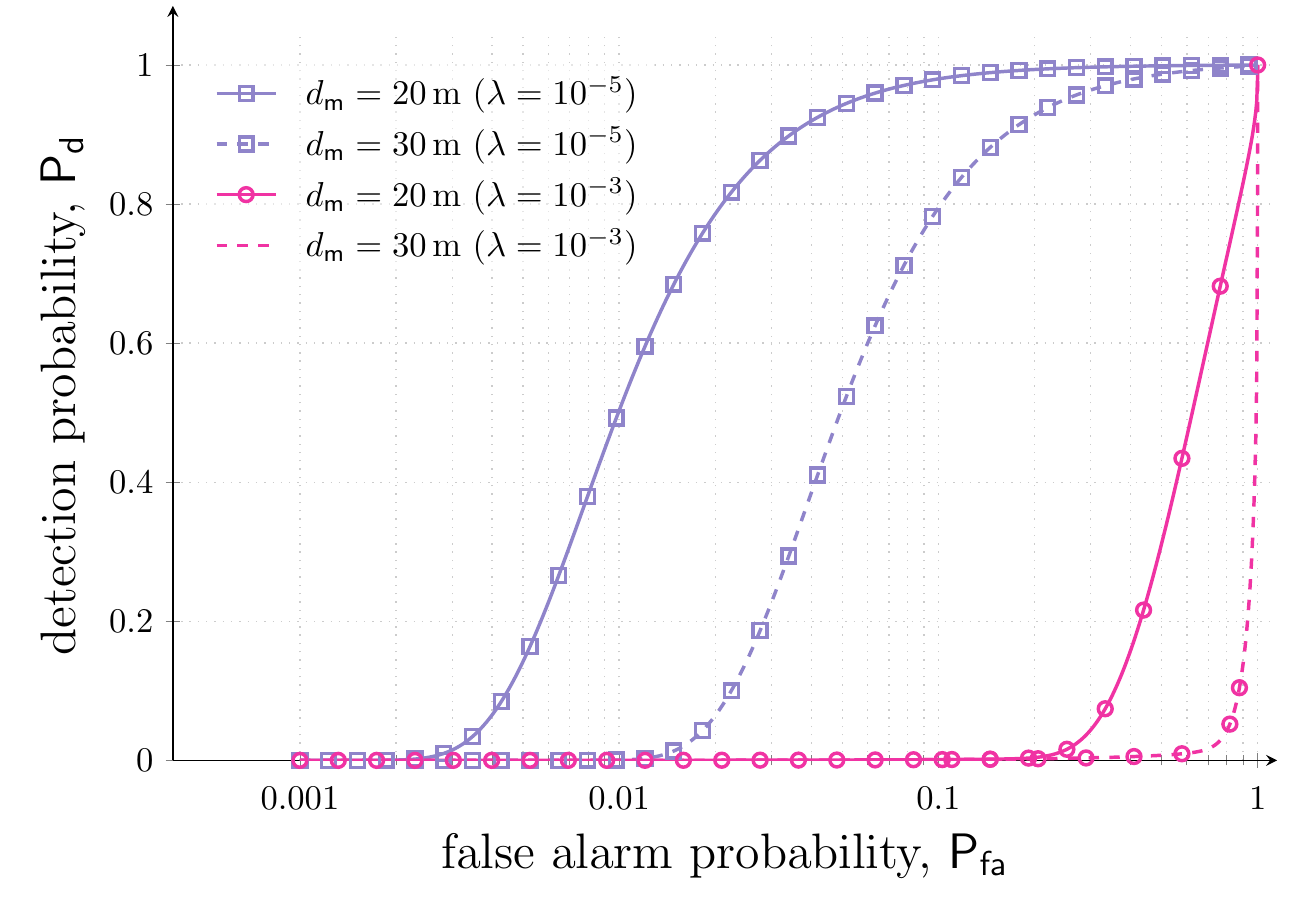}
\caption{Receiver Operating Characteristic (ROC) diagram, Rayleigh-fading case. Markers differentiate radar densities, while solid and dashed lines target distances. In all setups, $\varphi=\pi/6$, $\alpha=3$, \mbox{$f=2.4 \,\rm{GHz}$}.}
\label{fig:pdVsPfa_fading}
\vspace{-1em}
\end{figure}

Further insights on the system behaviour are offered by the ROC diagram in Fig.~\ref{fig:pdVsPfa_fading}, showing \pd\ vs. \pfa\ for different network densities. In sparse setups, e.g. $\lambda=10^{-5}$ [radar/m$^2$], a slight degradation of the detection performance can be effectively traded off for a lower \pfa, offering a useful design choice. By contrast, for larger $\lambda$, the plot emphasises that no reliable target identification is possible for the considered distances. This outcome highlights once more the strong role played by the transmitter density, strongly suggesting the need to design medium access strategies tailored for radar networks.

\appendices
\section{Impact of Noise on Detection}  \label{app:noise}

In order to gauge the effect of noise on detection performance, let us focus for simplicity on the no-fading case, and assume that a radar receiver is subject to a mixture of interference and random noise. Following a typical approach \cite{Nartasilpa16},  the latter can be modelled at the symbol level as a complex normal random variable with zero mean and variance $P_n = k_B T B F$, where $k_B$ is the Boltzmann constant, $T$ is the receiver operating temperature, $F$ is the receiver noise figure and $B$ is the bandwidth. In this case, detection is affected over a slot of interest by a non target-related power \mbox{$\mathcal Z := W+\mathcal I$}, where $W$ is an exponential r.v. of parameter $1/P_n$, i.e. \mbox{$W\sim \rm{exp}(1/P_n)$}, and $\mathcal I$ is the interference level captured by the stochastic geometry model presented in Sec.~\ref{sec:analysis}.  Observing that $W$ and $\mathcal I$ are statistically independent, and relying on the strongest interferer approximation for $\mathcal I$ (see Theorem~1), the cumulative distribution function of $\mathcal Z$ can be derived by means of convolution operations as:
\begin{align}
F_{\mathcal Z}(z) &:= \mathbb P\{  W + \mathcal I_s < z \} \nonumber\\
&= \frac{1}{P_n} \! \int_{0}^{z} \exp\left(  -\frac{\lambda \delta \varphi^2 \omega^{\frac{2}{\alpha}}}{4\pi} (z - w)^{-\frac{2}{\alpha}} - \frac{w}{P_n} \right) dw\,.
\label{eq:jointCDF}
\end{align}

The detection threshold $\Theta$ needed to guarantee a desired false alarm rate \pfa\ can then be derived numerically from \eqref{eq:jointCDF}, solving $1 - F_{\mathcal Z}(\Theta)^{M-1} = \pfa$. In turn, the maximum detectable range may be obtained by setting $\mathcal S = \Theta$ and solving the equation with respect to the target distance, leading to
\begin{equation}
d_{\mathsf m} = \left( \frac{\kappa \sigma P_t \mathcal G^2 \ell}{4\pi \Theta} \right)^{\frac{1}{2\alpha}} .
\label{eq:dmNoiseAndInt}
\end{equation}
The closed-form expression in \eqref{eq:dmNoiseAndInt} clarifies that, in the presence of noise, transmission power $P_t$, operating frequency $f$ and bandwidth $B$ (embedded in $\Theta$) start to play a role in determining system performance.
\begin{figure}
\centering
\includegraphics[width=.5\columnwidth]{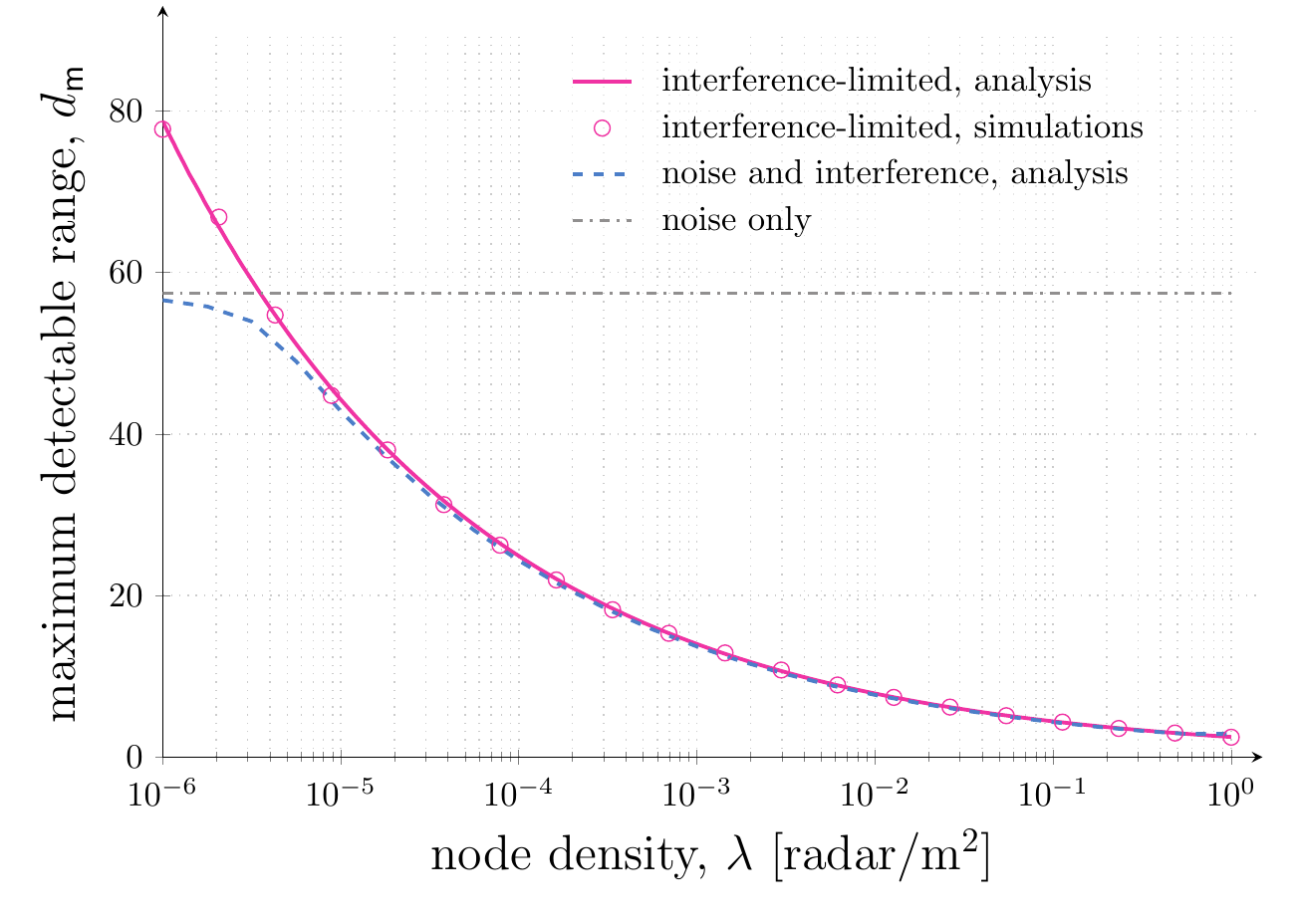}
\caption{\dm\ vs. radar density, no-fading case ($\alpha=2$, $f=60 \rm{GHz}$), with antenna beamwidth $\varphi=\pi/6$. The solid line shows results obtained assuming $\mathcal I \gg W$ and disregarding noise (Fig. $2$, p. $3$). Instead, the dashed blue line indicates the detectable range assuming a random noise component. The dash-dotted line shows performance with random noise, and in the absence of interference.}
\vspace{-3mm}
\label{fig:dmNoiseAndInt}
\end{figure}

Fig.~\ref{fig:dmNoiseAndInt} shows the trend of $d_{\mathsf m}$ against the network density $\lambda$, considering an antenna beamwidth $\varphi = \pi/6$, and setting, in the no-fading case under study ($\alpha=2$, $f = 60 \,\rm{GHz}$), $B = 125 \,\rm{MHz}$, $P_t = 20 \,\rm{dBm}$, $T=290 \,\rm{K}$, $F=10$. For convenience, the solid line replicates results obtained assuming an interference-limited system (i.e. disregarding noise) and already discussed in Fig.~\ref{fig:dmVsLambda}, whereas the dashed line is obtained assuming random noise as discussed above ($W\sim \exp(1/P_n)$).  As expected, when the device density increases, the effect of noise becomes negligible, and the two curves coincide. Conversely, for very low densities -- which are likely to not be of practical interest for most applications -- the detection behaviour becomes noise-limited. In this case, $d_{\mathsf m}$ saturates to a value that can easily be computed analytically. Namely, in the absence of interference, the detection threshold can be determined by setting \mbox{$\pfa = 1 - \mathbb P\{ W < \Theta \}^{M-1}$} to obtain
\begin{align}
\Theta = -P_n \ln \left( 1 - (1-\pfa)^{\frac{\delta}{1-\delta}} \right)
\end{align}
and, solving $\mathcal S = \Theta$ with respect to the target distance:
\begin{align}
d_{\mathsf m,\text{no-int}} = \left(  -\frac{P_t \mathcal G^2 \kappa \sigma \ell}{4 \pi P_n \ln \left( 1 - (1-\pfa)^{\frac{\delta}{1-\delta}} \right)} \right)^{\frac{1}{2\alpha}}
\end{align}
which is shown by the dash-dotted line in Fig.~\ref{fig:dmNoiseAndInt}. The results in Fig.~\ref{fig:dmNoiseAndInt} confirm the interference-limited nature of the networks under study for reasonable network densities, and underpin the ability of the presented approach to predict the fundamental trends and the effect of interference on radar detection performance.

\bibliographystyle{IEEEtran}
\bibliography{IEEEabrv,biblioRadar}

\end{document}